\documentclass{emulateapj}
\usepackage{apjfonts}
\usepackage{natbib}
\usepackage{epsfig}
\usepackage{amsmath}

\newcommand{\etal}{et al.}
\newcommand{\NH}{$N_{H}$}
\newcommand{\NHI}{$N_{H\,I}$}
\newcommand{\nh}{N_{H}}
\newcommand{\nhi}{N_{H\,I}}
\newcommand{\Mdot}{\dot{M}}
\newcommand{\Lbol}{L_{\rm bol}}
\newcommand{\dEdt}{\epsilon_r \Mdot c^{2}}

\newcommand{\dlgL}{d\log(L)}

\newcommand{\nLP}{\dot{n}(L_{peak})}

\newcommand{\meanNH}{\bar{N}_{H}}
\newcommand{\sigNH}{\sigma_{N_{H}}}

\shorttitle{Optical \&\ X-ray Quasar Luminosity Functions}
\shortauthors{Hopkins \etal}
\slugcomment{ApJ, Accepted, October 2005}

\begin{document}

\title{Luminosity-Dependent Quasar Lifetimes: \\
Reconciling the Optical and X-ray Quasar Luminosity Functions}
\author{Philip F. Hopkins\altaffilmark{1}, 
Lars Hernquist\altaffilmark{1}, 
Thomas J. Cox\altaffilmark{1}, 
Tiziana Di Matteo\altaffilmark{2}, 
Brant Robertson\altaffilmark{1}, 
Volker Springel\altaffilmark{3}}
\altaffiltext{1}{Harvard-Smithsonian Center for Astrophysics, 
60 Garden Street, Cambridge, MA 02138, USA}
\altaffiltext{2}{Carnegie Mellon University, 
Department of Physics, 5000 Forbes Ave., Pittsburgh, PA 15213}
\altaffiltext{3}{Max-Planck-Institut f\"{u}r Astrophysik, 
Karl-Schwarzchild-Stra\ss e 1, 85740 Garching bei M\"{u}nchen, Germany}

\begin{abstract}

We consider implications of our new model of quasar lifetimes and
light curves for the quasar luminosity function (LF) at different
frequencies and redshifts.  In our picture, quasars evolve rapidly and
the lifetime depends on both their instantaneous and peak luminosities.
The bright end of the LF traces the peak intrinsic
quasar activity, but the faint end consists of quasars which are
either undergoing exponential growth to much larger masses and
luminosities, or are in sub-Eddington quiescent states going into or
coming out of a period of peak activity.  The ``break'' in the
observed LF corresponds directly to the maximum in the intrinsic
distribution of peak luminosities, which falls off at both brighter
and fainter luminosities.  We study this model using simulations of
galaxy mergers which successfully reproduce a wide range of observed
quasar phenomena, including the observed column density distribution.
By combining quasar lifetimes and the distribution of maximum quasar
luminosities determined from the observed hard X-ray LF with the
corresponding luminosity and host-system dependent column densities,
we produce the expected soft X-ray and B-band LFs.  Our predictions
agree exceptionally well with the observed LFs at all observed
luminosities, over the redshift range considered ($z\leq1$),
without invoking any ad hoc assumptions about an obscured population
of sources. Our results also suggest that observed correlations in 
hard X-ray samples between 
the obscured fraction of quasars and luminosity can be explained 
in the context of our model by the expulsion of surrounding 
gas due to heating from 
accretion feedback energy as a quasar nears its peak luminosity and 
final black hole mass. 
\end{abstract}

\keywords{quasars: general --- galaxies: nuclei --- galaxies: active --- 
galaxies: evolution --- cosmology: theory}

\section{Introduction\label{sec:intro}}

The nature and evolution of the luminosity function (LF) of quasars at
different redshifts and frequencies has been studied for more than
thirty years \citep[e.g.,][and references
therein]{Schmidt68,SG83,Boyle00,Miyaji00,Ueda03}, but its relationship
to the intrinsic properties of individual quasars is not
well-understood.  Spectral synthesis modeling of the X-ray background
\citep[e.g.,][]{Comastri95,Gilli99,Gilli01} as well as observed differences
between hard X-ray and soft X-ray or optical quasar LFs
\citep[e.g.,][]{Boyle98,LaFranca02,Ueda03} imply (and require) a large
population of optically obscured quasars. However, unified models of
active galactic nuclei (AGN) \citep[e.g.,][]{Antonucci93} which invoke
geometric forms of obscuration as the dominant source of absorption
cannot predict the distribution of column densities or differences
between LFs, but rather depend on these observations to determine the
modeled form of obscuration. Even when calibrated by observed
ratios of obscured to unobscured AGN, such models cannot account
for measured quasar lifetimes or the selection-effect dependent
differences in observed LFs at different frequencies and redshifts.
Furthermore, a growing body of observations imply isotropic or
evolution-dependent obscuration which cannot be explained by these
simple models alone
\citep[e.g.,][]{Boronson92,Kuraszkiewicz00,Tran03,Page04,Barger05,
Alexander05,Stevens05}.

Previous efforts to interpret the quasar LF have relied on restrictive
assumptions about lifetimes and light curves of quasars, supposing,
for example, that quasars either have universal lifetimes or that they
evolve exponentially with time.  Semi-analytical modeling of the LF
\citep[e.g.,][]{KH00,Volonteri03,WL03} has neglected the obscured
quasar population and generally focused on reproducing the observed
optical or soft X-ray LF, which not only has a different shape but
also under-predicts the total quasar population by an order
of magnitude at most redshifts and luminosities.

Recently, we have begun to explore the impact of black hole growth on
galaxy formation, using simulations of galaxy mergers \citep{SDH05b}.
Our models reproduce the observed correlation between black hole mass
and galaxy velocity dispersion (the $M_{BH}-\sigma$ relation)
\citep{DSH05}, and link the quasar phase of galaxies \citep{H05a,H05b}
to galaxy evolution \citep{SDH05a}.  Furthermore, the simulations
predict qualitatively different quasar light curves than have been
adopted in previous work \citep{H05a,H05b}.  In our picture, the peak,
exponential black hole growth is determined by the gas supply over
timescales $\sim 10^{8}\,{\rm yr}$, during which the gas inflows
powering accretion generate large obscuring column densities.  The
growth shuts down when significant gas is expelled as it is heated by
feedback from black hole accretion, creating a window during which the
AGN is observable as an optical quasar for a lifetime
$\sim10^{7}\,$yr, in good agreement with observations, and yielding a
significant obscured quasar population \citep{H05a}.  \citet{H05b}
analyzed simulations over a range of galaxy masses and found that the
quasar light curves and lifetimes are all qualitatively similar, with
both the intrinsic and observed quasar lifetimes being strongly
decreasing functions of luminosity, with longer lifetimes at all
luminosities for higher-mass (higher peak luminosity) systems.
Moreover, they found that the resulting distribution of column
densities \NH\ depends significantly on the observed luminosity
threshold, and agrees remarkably well with observed \NH\ distributions
of both optical and X-ray samples once the appropriate selection
effects are applied.

In \citet{H05c} we discuss the intrinsic distribution of source
properties obtained by applying our model to the quasar LF,
recognizing the essential and realistic property that the time spent
at a given luminosity depends on both that luminosity and the peak
luminosity of the quasar (or, equivalently, the final black hole mass
or host system properties). This results in a qualitatively different
distribution of source properties than that implied by the idealized
light curves that have been used earlier.

Here, we combine our model of quasar lifetimes and the resulting
distribution of intrinsic source properties with the luminosity and
host system-dependent \NH\ distributions described above.  With these
self-consistent results derived from
hydrodynamical simulations, we find that the typical column density
distribution is a strong function of the instantaneous luminosity of a
quasar, and fit it to simple analytical functions. Using the observed
hard X-ray quasar LF to recover the distribution of intrinsic source
properties, we then combine our models of quasar lifetimes and the
corresponding observed column density distributions to reproduce the
expected LF at other frequencies given some absolute
magnitude/luminosity limit.  We find that our predictions for the
optical B-band and soft X-ray LFs agree well with
observations in both bands.  Thus, our model of quasar evolution,
without any assumptions, naturally reproduces differences in the hard
X-ray, B-band, and soft X-ray LFs over a range of
redshifts.

\section{The Simulations\label{sec:sim}}

The simulations presented here are the series described in detailed in \citet{H05b}, 
performed with
GADGET-2 (Springel 2005), a new version of the parallel TreeSPH code GADGET
\citep{SYW01} based on a fully conservative formulation
\citep{SH02} of smoothed particle hydrodynamics (SPH), which is
required for energy and entropy to be simultaneously conserved when
smoothing lengths evolve adaptively (see e.g., Hernquist 1993, O'Shea
et al. 2005).  Our simulations account for radiative cooling, heating
by a UV background (as in Katz et al. 1996b, Dav\'e et al. 1999), and
incorporate a sub-resolution model of a multiphase interstellar medium
(ISM) to describe star formation and supernova feedback \citep{SH03}.
Feedback from supernovae is captured in this sub-resolution model
through an effective equation of state for star-forming gas, enabling
us to stably evolve disks with arbitrary gas fractions (see,
e.g. Springel et al. 2005b; Robertson et al. 2004). 

Supermassive black holes (BHs) are represented computationally by
``sink'' particles that accrete gas at a rate $\Mdot$ estimated from
the local gas density and sound speed using an Eddington-limited
prescription based on Bondi-Hoyle-Lyttleton theory.  The
bolometric luminosity of the black hole is $\Lbol=\dEdt$, where
$\epsilon_r=0.1$ is the radiative efficiency.  We assume that a small
fraction (typical $\approx 5\%$) of $\Lbol$ couples dynamically to the
surrounding gas, and that this feedback is injected into the gas as
thermal energy.  This fraction is a free parameter, which we determine
as in \citet{DSH05} by matching the observed $M_{\rm BH}-\sigma$
relation.  For now, we do not resolve the small-scale dynamics of the
gas in the immediate vicinity on the black hole, but assume that the
time-averaged accretion rate can be estimated from the gas properties
on the scale of our spatial resolution ($\lesssim30$\,pc).

The progenitor galaxies in our merger simulations form  a family with 
virial velocities $V_{vir}=80, 113, 160, 226, {\rm
and}\ 320\,{\rm km\,s^{-1}}$.
The gas equation of state follows the multi-phase, star-forming structure 
derived in \citet{SH03}, resulting in a mass-weighted
temperature of star forming gas $\sim10^{5} {\rm K}$. 
For each simulation, we generate two stable, isolated disk galaxies,
each with an extended dark matter halo with a \citet{Hernquist90}
profile, motivated by cosmological simulations (e.g. Navarro et
al. 1996; Busha et al. 2004) and observations of halo properties
(e.g. Rines et al. 2000, 2002, 2003, 2004), an exponential disk of gas
and stars, and a bulge.  The self-similarity of any subset of these
models is broken by the scale-dependent physics of cooling, star
formation, and black hole accretion. The galaxies have masses $M_{\rm
vir}=V_{\rm vir}^{3}/(10GH_{0})$ with the baryonic disk
having a mass fraction $m_{\rm d}=0.041$, the bulge $m_{\rm b}=0.0136$,
and the rest of the mass in dark matter with a concentration parameter
$9.0$.  In \citet{H05a}, we describe our analysis of
simulation A3, one of our set with $V_{vir}=160\, {\rm km\,s^{-1}}$, 
a fiducial choice with a
rotation curve and mass similar to the Milky Way. We begin our
simulation with pure gas disks, which may better correspond to the
high-redshift galaxies in which most quasars are observed.

Each galaxy is initially composed of 168000 dark matter halo
particles, 8000 bulge particles, 24000 gas and 24000
stellar disk particles, and one BH particle.  We vary the initial seed
mass of the black hole to identify any systematic dependence of our
results on this choice.  In the cases considered, we choose the seed mass to be
sufficiently small that its presence will not have an immediate
effect.  Given the particle numbers employed, the dark matter, gas,
and star particles are all of roughly equal mass, and central cusps in
the dark matter and bulge profiles are reasonably well resolved (see
Fig 2. in Springel et al. 2005b).  The galaxies are then set to
collide from a zero energy prograde orbit.

\section{Column Densities in the Simulations\label{sec:NH}}

\subsection{Determining Column Densities to the Quasar\label{sec:NHmethods}}

We determine the column density between a black hole and a distant
observer as follows \citep{H05a,H05b}. 
We calculate the column density between a black hole and a
hypothetical observer from simulation outputs spaced every 10 Myr
before and after the merger and every 5 Myr during the merger of each
galaxy pair. We generate $\sim1000$ radial lines-of-sight (rays),
each with its origin at the black hole particle location and with
directions uniformly spaced in solid angle $d\cos{\theta}\,d\phi$. For
each ray, we then begin at the origin, calculate and record the local
gas properties using GADGET, and then move a distance along the ray
$\Delta r=\eta h_{\rm sml}$, where $\eta \leq 1$ and $h_{\rm sml}$ is
the local SPH smoothing length. The process is repeated until a ray is
sufficiently far from its origin ($\gtrsim 100$ kpc). The gas
properties along a given ray can then be integrated to give the
line-of-sight column density and mean metallicity.  We test different
values of $\eta$ and find that gas properties along a ray converge
rapidly and change smoothly for $\eta=0.5$ and smaller. We similarly
test different numbers of rays and find that the distribution of
line-of-sight properties converges for $\gtrsim 100$ rays.

Given the local gas properties, we use the GADGET multiphase model of
the ISM described in \citet{SH03} to calculate the local mass fraction
in ``hot'' (diffuse) and ``cold'' (molecular and HI cloud core) phases
of dense gas and, assuming pressure equilibrium between the two phases,
we obtain the local density of the hot and cold phase gas and the
corresponding volume filling factors. The values obtained correspond roughly
to the fiducial values of \citet{MO77}. A detailed description of the properties of 
both cold and hot phases of the resulting multiphase structure 
can be found in \citet{SH03}, Briefly, cold phase (HI or molecular) clouds, with a 
temperature $\lesssim1000\,$K, contain a fraction $\gtrsim0.9$ of 
the total gas mass above a critical star-forming density threshold determined by the 
equilibrium solutions to energy balance equations for injection by supernova feedback 
and radiative cooling. Given a temperature for the
warm, partially ionized component $\sim8000\,{\rm K}$ (with densities 
$n_{H,\,{\rm hot}}\sim0.1\,{\rm cm^{-3}}$, or more accurately 
$\sim0.01-0.1\,n_{H,\,{\rm total}}$), determined by
pressure equilibrium, we further calculate the neutral fraction of
this gas, typically $\sim0.3-0.5$.  We denote the neutral and total
column densities as \NHI\ and \NH, respectively. Using only the
hot-phase density allows us to place an effective lower limit on the
column density along a particular line of sight, as it assumes a ray
passes only through the diffuse ISM, with $\gtrsim 90\%$ of the mass
of the dense ISM concentrated in cold-phase ``clumps.'' Given the
small volume filling factor ($<0.01$) and cross section of such
clouds, we expect that the majority of sightlines will pass only
through the ``hot-phase'' component.

We assume the intrinsic quasar continuum SED follows
\citet{Marconi04}, based on optical through hard X-ray observations
\citep[e.g.,][]{Elvis94,George98,VB01,Perola02,Telfer02,Ueda03,VBS03}.
For the extinction at different frequencies, we consider a gas-to-dust
ratio equal to that of the Milky Way, $(A_{B}/\nhi)_{\rm
MW}=8.47\times10^{-22}\,{\rm cm^{2}}$, but scaled by metallicity,
$A_{B}/\nhi = (Z/0.02)(A_{B}/\nhi)_{\rm MW}$, as suggested by
observations \citep[e.g.,][]{Bouchet85}, although \citet{H05a} note
that the resulting difference is small.  We use the Small Magellanic
Cloud (SMC)-like reddening curve of \citet{Pei92}, again motivated by
observations \citep{Hopkins04}. We calculate extinction in X-ray
frequencies (0.03-10 keV) using the photoelectric absorption cross
sections of \citet{MM83} and non-relativistic Compton scattering cross
sections, similarly scaled by metallicity. In estimating the column
density for photoelectric X-ray absorption, we ignore the calculated
ionized fraction of the gas, as it is expected that the inner-shell
electrons which dominate the photoelectric absorption edges will be
unaffected in the temperature ranges of interest. We do not perform a
full radiative transfer calculation, and therefore do not model
scattering or re-processing of radiation by dust in the infrared.

\subsection{Column Density Evolution During Black Hole Growth\label{sec:NHevolgrow}}

For each simulation, we consider \NH\
values at all times with bolometric luminosity $L=\dEdt$ in some
logarithmic interval, weighted by the total time along all sightlines
a given \NH\ is observed.  This then gives us a binned distribution 
$P(\nh | L)\,d\log(\nh) \propto t(\log(\nh),\ t(\log(\nh) + d\log(\nh)))$. 
This distribution as a function of (observed) luminosity is shown in detail 
for our fiducial Milky Way-like simulation ($V_{vir}=160\,{\rm km\,s^{-1}}$)
in Figure~3 of \citet{H05b}, in which we demonstrate that the 
resulting column density distribution reproduces both the 
typical column density distribution of hard X-ray selected  
quasar samples \citep[e.g.,][]{Ueda03}, and that of optically selected 
samples. In particular, the agreement with the column density distribution 
from the SDSS as determined in \citet{Hopkins04} is very good, once similar selection effects are 
applied. 

\begin{figure}
    \centering
    \includegraphics[width=3.5in]{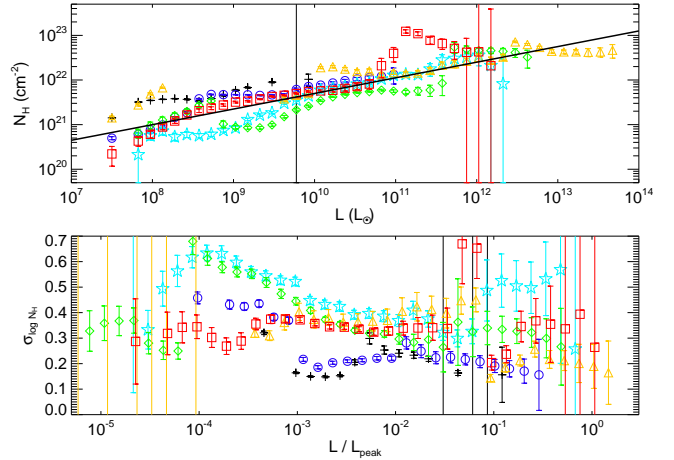}
    \caption{Lognormal median column density $\meanNH$ and dispersion $\sigNH$
    as a function of luminosity and the ratio of instantaneous to peak luminosity, for the 
    simulations described in \citet{H05b}. Points correspond to quasars with 
    final black hole masses of 
    $7\times10^{6}\, {\rm (black\ crosses)},\ 
      3\times10^{7}\, {\rm (blue\ circles)},\ 
      1\times10^{8}\, {\rm (red\ squares)},\       
      3\times10^{8}\, {\rm (cyan\ stars)},\ 
      7\times10^{8}\, {\rm (green\ diamonds)},\ {\rm and}\
      2\times10^{9}\,M_{\sun} {\rm (yellow\ triangles)}$. The black line 
      is the best-fit power law to $\meanNH(L)$.
    \label{fig:NH}}
\end{figure}

At each intrinsic (un-attenuated) bolometric luminosity $L$ we make a simple approximation
to the observed distribution and fit it to a lognormal form,
\begin{equation}
P(N_{H}) = \frac{1}{\sigNH \sqrt{2 \pi}}\ \exp [-(\log(N_{H}/\meanNH )^{2})/(2 \sigNH^{2})]. 
\end{equation}
We show the resulting total (neutral plus ionized) $\meanNH$ and $\sigNH$ for all simulations as a
function of $L$ in Figure~\ref{fig:NH}.  
Although we have shown this for the similar set of simulations described above, we 
have tested the column density distribution as a function of luminosity across a wide range 
of simulations, varying the gas fractions, orbital parameters, gas equations of state, 
concentrations, presence or 
absence of bulges, and seed black hole masses, and find that, although the final black hole mass 
(and corresponding peak luminosity) 
in any of these cases can be dramatically changed, the column density distribution as a 
function of instantaneous
and peak luminosity shows no systematic dependence on any of these 
host properties (Hopkins et al., in preparation). Therefore we can adopt these fits with reasonable 
confidence across a wide range of redshifts and luminosities. Indeed, just 
the simulated luminosities above range from $10^{7}-10^{14}\,L_{\sun}$, covering the 
entire range of actual {\em observed} quasar luminosities at almost all redshifts. 

We find that the dependence of $\sigNH$ on $L$ is weak, and we
consider both constant $\sigNH=0.4$ and a linear best-fit
$\sigNH=0.7+0.1\log{(L/L_{peak})}$. It is important to note, however, that this gives only the 
dispersion for an individual simulation; in calculating statistical quantities such as the luminosity function, 
the dispersion across a population is needed. This is easy to determine based on the dispersion in 
$\meanNH$ across simulations. Since we have assumed the distributions are individually lognormal, 
the dispersion of the quasar population is simply broadened to $\sigNH\approx0.8$ 
(or $\sigNH=1.0+0.1\log{(L/L_{peak})}$).
There is a clear trend of
increasing $\meanNH$ with $L$, which we fit to a power-law, giving
\begin{equation}
\meanNH\approx4.2\times10^{21}\,{\rm cm^{-2}}\,(L/10^{11}\,L_{\sun})^{0.35}.
\end{equation}
We note that the median neutral column density, $\bar{N}_{H\,I}$, follows a similar relation, 
with a typical ionized fraction $0.3-0.5$ (mean 0.35). 
This form can be understood roughly as follows, in the context of 
buried quasar growth during times when the black hole mass is growing to 
its final mass, before peak quasar stages. Consider the time-dependent
mass $M_{c}$ within the merging core of radius $R_{c}$ ($\sim100\,{\rm pc}$), 
and assume that the black hole grows such that $M_{BH}\sim\eta M_{c}$
($\eta\sim0.005$) \citep{mag98}. The total density is then $\rho=M_{c}
R_{c}^{-3} f(r)$, where $f(r)$ is a dimensionless profile of order
unity, and the column density is $\nhi=(f_{\ast} M_{c}) / (\mu m_{H}
R_{c}^{2})$, where $f_{\ast}$ is the product of the mean neutral,
hot-phase fraction ($\lesssim0.01$ in the most dense regions of the
galaxy) and the integral of the density profile $f(r)$ ($\sim1$), and
$\mu$ is the mean molecular weight. This gives
$\nhi\approx10^{21}\,{\rm cm^{-2}}\,m_{8}$, where
$M_{c}=m_{8}\times10^{8}\,M_{\sun}$.
The luminosity, from a Bondi accretion model, is 
$L=\epsilon_{r}\dot{M}c^{2}=4\pi\alpha\epsilon_{r}(G M_{BH})^{2}\rho c^{2}/c_{s}^{3}$, 
or using the definitions above, 
$L=(4\pi\alpha\epsilon_{r}\eta^{2} f_{0}) (c/c_{s})^{3} (G^{2} M_{c}^{3}) / (c R_{c}^{3})$, 
where $f_{0}=f(r)\sim1$ as appropriate for some ``accretion radius'' and
$c_{s}$ is the (mass-weighted) sound speed ($\sim30\,{\rm km/s}$ in the approximately
isothermal core). These values give
$L\approx10^{11}\,L_{\sun}\,m_{8}^{3}$. Thus, we expect
$\nhi\sim10^{21}\,{\rm cm^{-2}}\,(L/10^{11}\,L_{\sun})^{1/3}$, 
with variation of the above parameters as well as dynamical and feedback
effects generating the considerable scatter seen in this 
relationship.  This relation is much shallower than the naively 
expected relation $\nh\propto L$ expected if $M_{BH}$ is 
constant ($L\propto\rho\propto\nh$) or $L\propto M_{BH}$ always, 
and strongly contrasts with unification models which predict static 
obscuration or  \NH\ independent of $L$ up to some threshold
\citep[e.g.,][]{Fabian99}.

This modeling naturally produces a population of heavily attenuated objects at 
large luminosities. Although our fitted hot-phase column densities do not 
reach extremely Compton-thick levels $\nh\gtrsim10^{25-26}\,{\rm cm^{-2}}$, 
the brightest objects shown in Figure~\ref{fig:NH} 
reach median column densities $\sim10^{23}\,{\rm cm^{-2}}$ or larger. 
As the fitted distribution is a lognormal about this median with a dispersion 
across the quasar population $\sigNH\sim1$, this implies a significant population of 
heavily attenuated ($\nh\sim10^{23}-10^{24}\,{\rm cm^{-2}}$) 
objects at soft X-ray (0.5-2 keV) frequencies; with the number density of sources 
falling of with larger \NH. This agrees well with both direct 
observations \citep{Treister04,Mainieri05} as well as synthesis models of 
the X-ray background \citep{Madau94,Comastri95,Gilli99,Gilli01}, 
which require a population of such objects, 
suggesting that our model should in principle account for the X-ray background spectrum 
based on a given luminosity function, 
without adopting arbitrary distributions of source obscuration or additional obscured 
populations. Furthermore, as discussed in \citet{H05b}, extension of this 
distribution of $\nh(L)$ to very bright quasars in unusually massive galaxies or
quasars in higher-redshift, compact galaxies which we have not
simulated may, during peak accretion periods, reach 
Compton-thick values ($\nh\gtrsim10^{25}\,{\rm cm^{-2}}$) 
of the typical column density. More likely, as our model assumes
$\sim90\%$ of the mass of the densest gas is clumped into cold-phase
molecular clouds, a small fraction of sightlines will pass through
such clouds and encounter column densities similar to those shown for the
cold phase in Figure~2 of \citet{H05a} and Figures~4 and 5 of \citet{H05b}, 
as large as $\nh\sim10^{26}{\rm cm^{-2}}$. This also allows a large concentration of mass in
sub-resolution obscuring structures, such as an obscuring toroid on
scales $\lesssim100\,$pc, although many of the phenomena such
structures are invoked to explain can be accounted for through our
model of time-dependent obscuration. Geometrical effects may therefore, however,  
become relevant for interpreting and explaining the Compton-thick 
source population. 

\subsection{Column Densities in Final Growth Stages \label{sec:NHfinal}}

Although the physical motivation for such a dependence is intuitive, 
the trend of increasing column density with luminosity seems to 
run opposite to that observed in a number of X-ray samples 
\citep{Steffen03,Ueda03,Hasinger04,GRW04,sazrev04,Barger05,Simpson05}.
In these samples, it appears that the fraction of broad-line objects and the 
fraction below some moderate column density ($\nh=10^{22}\,{\rm cm^{-2}}$) 
increase with luminosity, at odds with our prediction. However, a more detailed 
inspection reveals that our simulations 
can and do, in fact, reproduce this behavior. Although it is possible to extend our  
spectral modeling of the quasar and column density calculation to 
describe the complete stellar population of the galaxy, and so determine 
more accurately when an X-ray selected quasar will show a typical broad-line 
spectrum, we defer this to a later paper as it requires 
calculating 
colors and attenuations of the stars at all times in our simulations. 
However, we can easily compare to the observations of e.g., Ueda et al. (2003),
and consider the fraction of objects with $\nh<10^{22}\,{\rm cm^{-2}}$ 
(more accurately, the ratio of number with $\nh<10^{22}\,{\rm cm^{-2}}$ 
to all with $\nh<10^{24}\,{\rm cm^{-2}}$) as a function of hard X-ray luminosity. 

\begin{figure}
    \centering
    \includegraphics[width=3.5in]{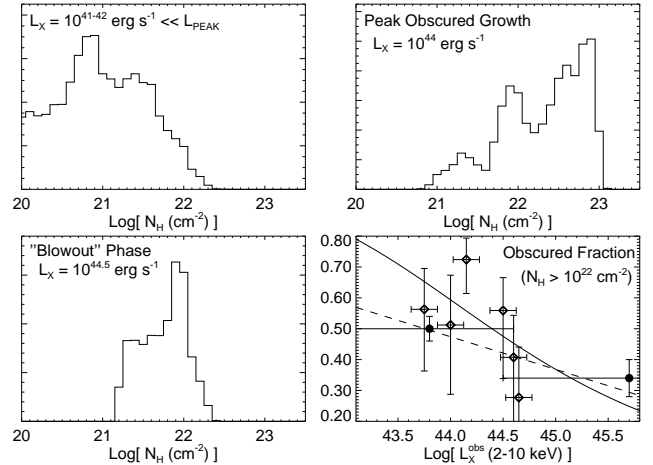}
    \caption{The column density distribution at three observed hard X-ray (2-10 keV) luminosities, from 
    the Milky Way like ($V_{vir}=160\,{\rm km\,s^{-1}}$) simulation of \citet{H05a}, 
    with a peak hard X-ray luminosity $L_{X,\ {\rm peak}}=5\times10^{44}\,{\rm erg\,s^{-1}}$. 
    The luminosities are chosen to be well below the peak luminosity (upper left) and near peak, corresponding
    to the final stages of obscured black hole growth (upper right) and subsequent ``blowout'' phase (lower left) 
    as feedback from accretion energy expels gas in the central regions of the galaxy. Lower right shows the 
    fraction of sightlines with $\nh>10^{22}\,{\rm cm^{-2}}$ from the six simulation outputs closest in 
    time and luminosity to $L_{X,\ {\rm peak}}$ (black diamonds), with horizontal errors the spacing in luminosity and 
    vertical errors corresponding to the range across different possible column density prescriptions 
    compared in \citet{H05b}. Solid line is the best-fit obscured fraction as a function of 
    luminosity from \citet{Simpson05}. Circles show the observations from \citet{Ueda03}, with the 
    dashed line the expectation from their fitted $\nh$ function. 
    \label{fig:compareUeda}}
\end{figure}

We examine the \NH\ distribution of our fiducial, Milky Way-like ($V_{vir}=160\,{\rm km\,s^{-1}}$)
simulation in detail in Figure~\ref{fig:compareUeda}. Over most of the simulation, 
we find the general trend shown in Figure~\ref{fig:NH} and discussed above. 
The upper left panel shows the binned column density distribution (arbitrary scale) 
for all times with an observed hard X-ray luminosity 
$L^{\rm obs}_{X}(2-10\, {\rm keV})=10^{41}-10^{42}\,{\rm erg\,s^{-1}}$ (we use 
these units for ease of comparison with the observations quoted above), well below the peak 
observed hard X-ray quasar luminosity $\approx5\times10^{44}\,{\rm erg\,s^{-1}}$, at 
which point the evolution is following the normal trend shown in Figure~\ref{fig:NH}. 
However, when the quasar nears its final, peak luminosity, there is a rapid ``blowout'' phase
as thermal feedback from the growing accretion heats the surrounding gas, driving a strong 
wind and eventually cutting off the accretion process, leaving a remnant with a black hole 
satisfying the $M_{\rm BH}-\sigma$ relation \citep{DSH05,H05a}. We analyze this ``blowout'' 
phase in some detail in \citet{H05a}, and find that it can be identified with the traditional 
bright optical quasar phase, as the final stage of black hole growth with a rapidly declining 
density (allowing the quasar to be observed in optical samples), giving 
typical luminosities, column densities, and lifetimes of optical quasars \citep{H05a,H05b} in 
good agreement with observations \citep[e.g.,][]{Hopkins04,Martini04}. 

If we consider these stages of 
quasar growth, then, near the peak luminosity, we find a very different trend. The upper right 
panel of Figure~\ref{fig:compareUeda} shows the column density distribution 
for times with $L^{\rm obs}_{X}\approx10^{44}\,{\rm erg\,s^{-1}}$, 
primarily in the final stages of 
the rapid obscured black hole growth phase, just before feedback from the accretion begins to 
expel the surrounding gas. The typical column densities are large, $\gtrsim10^{22}\,{\rm cm^{-2}}$, 
as we expect from our modeling above, although not high enough to 
generate large extinction in hard X-rays. However, just $\sim10$\,Myr later, the quasar has 
expelled a significant amount of gas and column densities fall rapidly. The lower left panel of 
the figure shows the column density distribution for times with a typical luminosity 
$L^{\rm obs}_{X}\approx10^{44.5}\,{\rm erg\,s^{-1}}$, essentially the very peak 
quasar luminosity, corresponding to the bright observable phase in which the quasar is driving 
a wind and expelling nearby gas. The typical column densities are lower by an order of 
magnitude, $\nh\sim10^{21}-10^{22}\,{\rm cm^{-2}}$. In the lower right 
panel of the figure, we plot the fraction of sightlines with column densities above 
$10^{22}\,{\rm cm^{-2}}$ (the ``obscured fraction''), for the six simulation outputs closest (in both 
time and luminosity) to the peak luminosity of 
the quasar. The results are shown (black diamonds) with the range in luminosity between each 
simulation output shown as horizontal error bars. In \citet{H05b} we 
discuss in detail the differences in column densities that result from varying our method of 
calculation, and find that (after accounting for some hot phase-cold phase separation) 
adopting various extreme cases yields a factor $\sim2$ difference in the calculated column densities, 
so we plot the resulting differences in the obscured fraction from this uncertainty as 
vertical error bars. The data from \citet{Ueda03} (circles) and their expectation from their 
fitted column density distribution (dashed lines) are shown for comparison, as is the 
best-fit obscured fraction as a function of luminosity from \citet{Simpson05} (solid line).  

The agreement between the observations and our result is encouraging, especially as our calculation 
considers only one simulation, and is not necessarily meant to 
reproduce the trend in observed quasar populations. However, we do find a similar trend in all 
of our simulations near the peak luminosity \citep[see also][]{H05a,H05b}. 
The key point is that we find, near the 
{\em peak} luminosity of the quasar as feedback drives away gas and shuts down accretion 
processes, the typical column densities fall rapidly 
with luminosity in a manner similar to that observed. In our model for the luminosity 
function, proposed in \citet{H05c}, quasars below the ``break'' in the observed 
luminosity function are either growing efficiently in early stages of growth or in sub-Eddington 
phases coming into or our of their peak quasar activity. Around and above the break in the luminosity 
function, the observed luminosity function becomes dominated by sources at high Eddington ratio 
at or near their peak luminosities. Based on the above calculation, as well as the 
description of the blowout phase from \citet{H05a}, we then immediately {\em expect} what is 
observed, that in this range of luminosities, the fraction of objects observed with large column densities 
will rapidly decrease with luminosity as the observed sample is increasingly dominated by sources at their 
peak luminosities in this blowout phase. This also further emphasizes that the evolution 
of quasars dominates over static geometrical effects in determining the observed 
column density distribution at any given luminosity.

\section{Quasar Lifetimes \&\ the Luminosity Function\label{sec:lifetimes}}

\citet{H05c} showed that a proper accounting of realistic quasar
light curves results in luminosity-dependent quasar lifetimes.  In
this picture, quasar lifetimes are functions of {\it both} the
instantaneous luminosity {\it and} the peak luminosity (i.e. final
black hole mass or host galaxy properties) of the system.  Given a
quasar lifetime above some luminosity as a function of the peak
luminosity of the quasar, $t(L'>L, L_{peak})$, the quasar LF
(in the absence of selection effects) is given by
\begin{equation}
\Phi(L)=\int{\frac{dt(L, L_{peak})}{\dlgL}\,\nLP}\,d\log(L_{peak}),
\label{eqn:phi0}
\end{equation}
where $\nLP$ is the rate at which of sources in a given logarithmic interval in $L_{peak}$ are 
``born'' (created or activated) per unit volume and
$\Phi(L)$ is the number density of sources per logarithmic
interval in $L$.  This formulation implicitly accounts for the ``duty
cycle'' (the fraction of active quasars at a given time), which
is proportional to the lifetime at a given luminosity.
At any redshift, $\nLP$ will be, in general, a complicated function of 
the distribution of galaxy properties, including merger rates, masses, and gas fractions. 
However, having determined the quasar lifetime from our simulations, we 
can use an observed luminosity function to de-convolve $\nLP$. Since we 
are only interested here in demonstrating that our modeling self-consistently 
reproduces the observed differences in luminosity functions, in concert with the 
interpretation of the luminosity function from \citet{H05c}, we adopt this 
semi-empirical approach in what follows, and thus all quantities in the equations 
determining the intrinsic and observed luminosity functions are completely determined. 

Given a distribution of \NH\ values and some minimum observed
luminosity $L_{\nu}^{min}$, the fraction $f_{obs}$ of quasars with a 
peak luminosity $L_{peak}$ and instantaneous bolometric 
luminosity $L$ which 
lie above the luminosity threshold is given by the 
fraction of \NH\ values below a critical $N_{H}^{max}$, where 
$L_{\nu}^{min}=f_{\nu}L\,\exp{(-\sigma_{\nu}N_{H}^{max})}$. Here,
$f_{\nu}(L)\equiv L_{\nu}/L$ is a bolometric correction and $\sigma_{\nu}$ 
is the cross-section at frequency $\nu$. Thus, 
\begin{equation}
N_{H}^{max}(\nu,L,L_{\nu}^{min})=\frac{1}{\sigma_{\nu}}\ln{\Bigl( \frac{f_{\nu}(L)L}{L_{\nu}^{min}}\Bigr)},
\end{equation}
and for the lognormal distribution above, 
\begin{equation}
f_{obs}(\nu,L,L_{peak},L_{\nu}^{min})=\frac{1}{2} \Bigl[ 1 + {\rm erf}\Bigl( \frac{\log{(N_{H}^{max}/\meanNH)}}{\sqrt{2}\,\sigNH}\Bigr)\Bigr].
\end{equation}
This results in a LF (in terms of the bolometric luminosity)
\begin{equation}
  \begin{split}
     \Phi(\nu,L,L_{\nu}^{min})& =\frac{1}{t_{\ast}}\int{f_{obs}(\nu,L,L_{peak},L_{\nu}^{min})} \\
     &\quad \times\frac{dt(L, L_{peak})}{\dlgL}\,\nLP\,d\log(L_{peak}).
     \label{eqn:phifull}
  \end{split}
\end{equation}
The important point to recognize in this equation is that $f_{obs}$ has a complicated 
dependence on both instantaneous and peak luminosity (which does not simply factor out of 
the above integrals), and thus the equations above will 
have a non-trivial dependence on the distribution of quasar peak luminosities $\nLP$ and 
the quasar lifetime, which are very different in our analysis from what has 
generally been used in previous modeling attempts. 

We consider quasar lifetimes 
determined from the simulations
described in \citet{H05a,H05b}. The light curves in the
mergers are complicated, generally having a period of early rapid accretion
after ``first passage'' of the galaxies, followed
by an extended quiescent period, then a transition to a peak, highly
luminous quasar phase, and then a dimming as self-regulated mechanisms
expel gas from the galaxy center after the black hole reaches a
critical mass and shut down accretion \citep{DSH05}.  While complex,
\citet{H05b} find that the total
quasar lifetime $t_{Q}(L'>L)$ above a given luminosity $L$ is 
well-approximated by a truncated power law for every simulation studied, with
$t_{Q}(L'>L) = t_{9}\,(L / 10^{9}\,L_{\sun})^{\alpha}$, 
where $t_{9}\equiv t_{Q}(L'>10^{9}\,L_{\sun})\sim10^{9}\,{\rm yr}$,
over the range $10^{9}\,L_{\sun}<L<L_{peak}$ for a given
quasar. Given then that \citet{H05b} find 
this normalization is approximately constant across simulations, the 
lifetime in each simulation is then entirely determined by 
the power-law slope $\alpha$. The value of $\alpha$ depends 
on the {\em peak} luminosity of the quasar (equivalently, the final black hole 
mass or host system properties), with more massive (higher peak luminosity) 
quasars yielding shallower power-law slopes as they spend more time at 
high luminosities and large Eddington ratios. 
Over a wide range of $L_{peak}$ (from
$\sim10^{10}-10^{14}\,L_{\sun}$), \citet{H05b} find 
$\alpha=\alpha(L_{peak})$ is approximately linear with $\log{L_{peak}}$, 
$\alpha=\alpha_{0}+\alpha'\,\log{L_{peak}}$ with an upper limit 
$\alpha=-0.2$.
The time spent in any logarithmic luminosity
interval in this range is then simply
\begin{equation}
dt/\dlgL = |\alpha|\,t_{9}\,(L / 10^{9}\,L_{\sun})^{\alpha}. 
\end{equation}
\citet{H05c} examine a series of possible restrictions to this model, which 
change $\alpha(L_{peak})$ slightly but yield very
similar behavior, and we have considered 
all the cases described therein and obtain 
identical results in every case using luminosity-dependent 
quasar lifetimes of this form. We show results for the case in which 
$\alpha$ is determined from the entire duration of our simulations, without imposing 
any arbitrary cutoffs, which gives $\alpha_{0}=-0.95$ and $\alpha'=0.32$ for 
$L_{peak}$ in units of $10^{12}\,L_{\sun}$, i.e.\ 
$\alpha=-0.95 + 0.32\log{(L_{peak}/10^{12}L_{\sun})}$. 

Given this model of $dt/\dlgL$ as a function of $L$ and $L_{peak}$, we
can then fit to any $\Phi(L)$ to determine $\nLP$. The resulting
distributions $\nLP$, discussed in \citet{H05c} are fundamentally
different from the naive expectation of previous analyses of the
LF which relied on idealized 
models of the quasar lifetime, either assuming quasars ``turn on'' at
a fixed luminosity for some universal lifetime,
($dt/\dlgL\propto\delta(L-L_{0})$) or assuming a pure exponential
light curve over some interval ($dt/\dlgL={\rm constant}$).
These previous models for the quasar lifetime 
yield a direct relationship between an observed luminosity and 
peak luminosity (final black hole mass), and predict a distribution 
of peak luminosities $\nLP$ with essentially identical shape to the 
observed LF. However, accounting for the luminosity 
dependence of quasar lifetimes based on our detailed modeling, we find that 
quasars spend significantly more time at low luminosities than near their peak, resulting 
in a very different $\nLP$ distribution. In this modeling, $\nLP$ has the same shape 
as the observed LF above the ``break'' in the luminosity function, and these quasars 
are accreting at high efficiency, near their peak luminosities. Below the break luminosity, 
however, the $\nLP$ distribution turns over, and the faint end of the luminosity function 
is dominated by sources near the break luminosity (the {\em peak} of the $\nLP$ distribution) 
accreting either at high efficiency but in early stages of merger and growing to much larger
luminosities, or in sub-Eddington phases going into or out of peak quasar activity (in the 
final stages of the host galaxy merger). The evolution of the LF with redshift, then, directly 
relates to evolution in $\nLP$, with the characteristic peak luminosity of quasars (final black 
hole mass being built) increasing with redshift as the break luminosity shifts to larger values. 
It is of particular interest to determine if this modeling of quasar evolution and the consequent 
novel interpretation of the luminosity function produce differences in the luminosity function 
in different wavebands (as well as column density distributions) consistent with what is observed. 
Furthermore, when comparing different models of quasar lifetimes and the luminosity function, 
even for $\nLP$ distributions chosen such that two different quasar lifetime models will produce 
an identical luminosity function in a given waveband, 
any dependence of the column density distribution on peak luminosity (i.e.\ host system properties) 
or Eddington ratio will result in a different prediction for the luminosity function at all other 
frequencies. 

\begin{figure}
    \centering
    \includegraphics[width=3.5in]{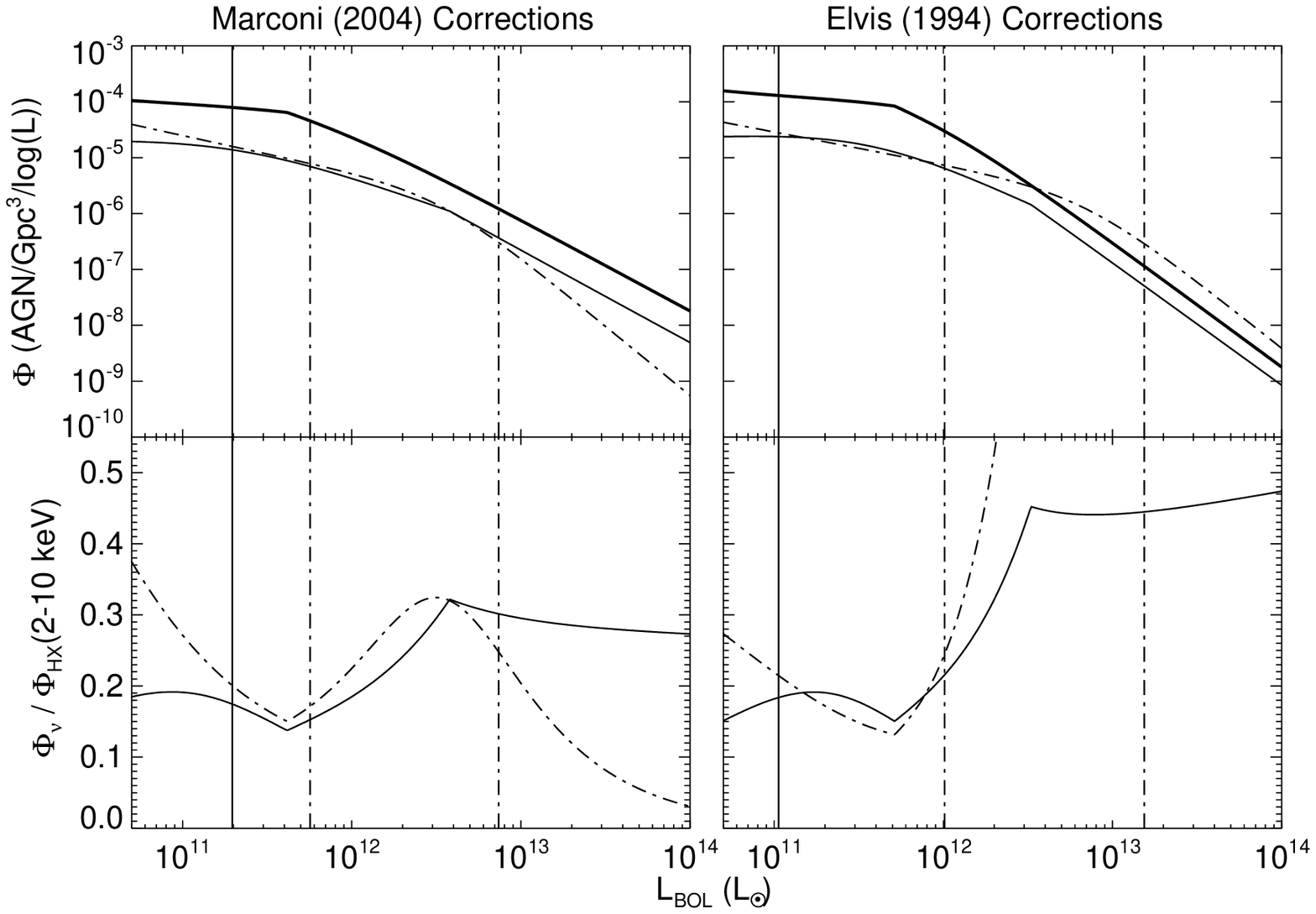}
    \caption{Upper panels: Hard X-ray \citep[thick;][]{Ueda03}, soft X-ray \citep[thin;][]{Miyaji00}, 
    and B-band \citep[dot-dash;][]{Boyle00} LFs at $z=1.0$, rescaled using the bolometric corrections from 
    \citet{Marconi04} which we adopt throughout (left) and the luminosity-independent average 
    corrections from \citet{Elvis94} (right). Lower panels: The ratio of soft X-ray (solid) and 
    B-band (dot-dash) LFs to the hard X-ray LF above. 
    LFs are plotted over the corresponding range of the observed luminosities in 
    the \citet{Ueda03} hard X-ray LF at this redshift.  
    Vertical lines bracket the minimum and maximum bolometric luminosities for which 
    observational data exist (according to the appropriate bolometric correction) in the soft 
    X-ray (solid line, only a minimum shown) and B-band (dot-dashed lines). Outside of these ranges 
    the fitted luminosity functions are extrapolated beyond the data. 
    \label{fig:bolcor}}
\end{figure}

We consider the observed luminosity functions in the hard X-ray ($\Phi_{HX}$; 2-10 keV), 
soft X-ray ($\Phi_{SX}$; 0.5-2 keV), and optical B-band  ($\Phi_{B};\
\lambda_{B}=4400\,$\AA), from \citet{Ueda03,Miyaji00,Boyle00}, respectively. 
We rescale all luminosity functions to the same $\Omega_{m}=0.3$, 
$\Omega_{\Lambda}=0.7$, $H_{0}=70\,{\rm km\,s^{-1}\,Mpc^{-1}}$ cosmology. 
In order to make a direct comparison between luminosity functions, 
we further rescale all luminosity functions in terms of the bolometric luminosity, 
using the bolometric corrections of \citet{Marconi04}. 
It is important to note that using the constant, luminosity-independent 
bolometric corrections of e.g.\ \citet{Elvis94} 
instead results in a significantly steeper cutoff in the luminosity function at high bolometric 
luminosities, as the bolometric luminosity inferred for the brightest observed X-ray quasars is almost 
an order of magnitude smaller using the \citet{Elvis94} corrections. However, 
it has been well-established that 
the ratio of bolometric luminosity to hard or soft X-ray luminosity increases with increasing 
luminosity \citep[e.g.,][]{Wilkes94,Green95,VBS03,Strateva05}, and further the 
sample quasars of \citet{Elvis94} are X-ray bright \citep{ERZ02}. 
Accounting for the luminosity dependence of the UV to X-ray flux ratio, 
$\alpha_{OX}$, gives rise to most of this difference. We adopt 
the form for $\alpha_{OX}$ from \citet{VBS03}, but our 
results are relatively insensitive to the different values found in the literature. 
These differences in the resulting bolometric luminosity functions are illustrated in 
Figure~\ref{fig:bolcor}. The upper left shows the luminosity functions 
$\Phi_{HX}$ (thick), $\Phi_{SX}$ (thin), and $\Phi_{B}$ (dashed) at $z=1.0$, with 
bolometric luminosities from the \citet{Marconi04} corrections and 
corresponding densities rescaled according to 
\begin{equation}
\frac{d\Phi}{d\log{L}} = \frac{d\Phi}{d\log{L_{\nu}}}\frac{d\log{L_{\nu}}}{d\log{L}}.
\end{equation}
This can be compared to the same luminosity functions converted using the 
constant bolometric corrections of \citet{Elvis94} (upper right). In both 
cases the qualitative differences are similar, with the hard X-ray luminosity function 
significantly above the soft X-ray or optical LF at low luminosities just below the break, 
and the ratio of hard X-ray to soft X-ray or optical LF decreasing at and just above the break. 
This is seen in the lower panels of the figure, where we plot the corresponding ratios 
$\Phi_{SX}/\Phi_{HX}$ (solid) and $\Phi_{B}/\Phi_{HX}$ (dot-dash).
However, it is apparent that the inferred number density from the X-ray LFs decreases much 
more slowly with luminosity using the \citet{Marconi04} corrections which account 
for the luminosity dependence of $\alpha_{OX}$, resulting in a larger gap at a given 
luminosity between the hard and soft X-ray LFs (and the optical as well). 
It is also immediately clear in this plot that the constant bolometric corrections of 
\citet{Elvis94} cannot apply uniformly to all luminosities and redshifts, as this 
actually predicts a {\em larger} number of optically selected bright quasars than 
soft or hard X-ray objects, which cannot be explained with any sort of 
reddening/obscuration model. This is not a fitting artifact, as a direct comparison of the 
data between, e.g., \citet{Croom04} and \citet{Barger05} in the optical and 
hard X-ray, respectively, shows (using the \citet{Elvis94} corrections) the optical 
quasar LF to be a factor $\sim2-3$ higher than the hard X-ray LF at several 
luminosities and redshifts, and this problem is only worse when considering the 
optical LFs of \citet{Boyle00} or \citet{Richards05} which are steeper at low luminosity 
than that of \citet{Croom04}. 
Furthermore, there is direct evidence of a significant number of optically obscured 
quasars, even at high luminosities 
\citep{Norman02,Stern02,Dawson03,Treister04}, many of which are undergoing 
a buried quasar or pre-quasar growth phase 
\citep{Page04,Alexander05,Stevens05} as predicted in our modeling. 
As discussed above, the existence of 
a significant population of such objects is also inferred from synthesis models of 
the X-ray background \citep{Madau94,Comastri95,Gilli99,Gilli01}. 
At the extreme low and high luminosities shown, the relative behavior of the 
plotted LFs becomes confusing, but this is because they are extrapolated well beyond 
the range of observed data. Therefore, we show the actual range of observed luminosities 
for each LF as vertical lines (of the same style as the corresponding LF). The difference 
in LFs across these ranges are much less dramatic, and this is what we are interested 
in making a detailed comparison with. 
It is clear, though, that it is important to account for the luminosity dependence of quasar bolometric corrections, 
as it creates this significant difference in the high-luminosity end of the 
bolometric quasar luminosity function and implies that a non-negligible fraction of the brightest quasars 
are not seen in optical surveys, and further that the ratio of one luminosity function to another is 
not trivially related to the obscured fraction discussed in detail above. 
For further comparison of these bolometric corrections we refer to \citet{Marconi04}, 
and for further detailed comparison of luminosity functions using a very similar 
procedure to produce the bolometric corrections see, e.g., \citet{Richards05}.

\begin{figure}
    \centering
    \includegraphics[width=3.5in]{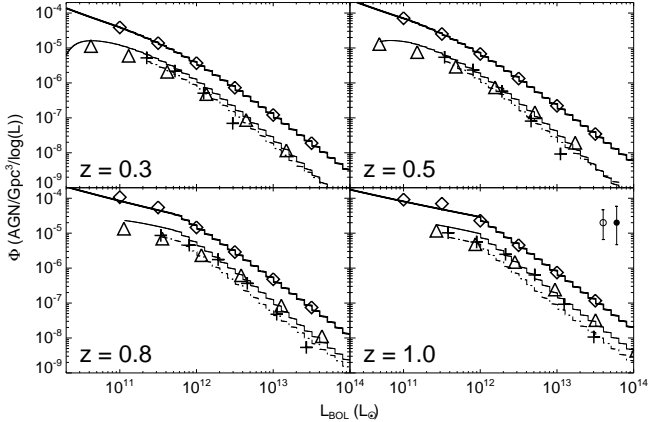}
    \caption{Hard X-ray (thick), soft X-ray (thin), and B-band (dot-dash) LFs 
    determined from our model of quasar lifetimes and column densities, based on 
    a distribution of intrinsic source properties fitted to the observed hard X-ray LF
    as in \citet{H05c} 
    and the limiting magnitudes of observed samples, at the different redshifts shown. 
    All quantities are rescaled to bolometric luminosities with the bolometric corrections of \citet{Marconi04}.
    Symbols show the observed LFs over the range where observations exist, 
    for hard X-rays \citep[][diamonds]{Ueda03}, soft X-rays \citep[][triangles]{Miyaji00}, 
    and B-band \citep[][crosses]{Boyle00}. In the $z=1.0$ panel, the points at upper right 
    show the mean systematic offset in the soft X-ray (open) and B-band (filled) luminosity 
    functions which result from modifying our column density prescription to ignore ionization, metallicity, 
    or vary the attenuation calculation following the range of possibilities studied in \citet{H05b}.
    \label{fig:LF}}
\end{figure}

Figure~\ref{fig:LF} shows the resulting LFs in
different bands at redshifts $z=0.3,0.5,0.8,\ {\rm and }\ 1.0$.  
All quantities have been rescaled in terms of the
bolometric luminosity as discussed above. 
We calculate the $\nLP$ distribution by fitting to
the observed hard X-ray (2-10 keV) LF of
\citet{Ueda03}, $\Phi_{HX}$ at each redshift. Given this $\nLP$, we then use our model
of quasar lifetimes and the \NH\ distributions determined in \S2\ 
to calculate the expected B-band ($\Phi_{B};\
\lambda_{B}=4400\,$\AA) and soft X-ray ($\Phi_{SX}$; 0.5-2 keV)
LFs, given the appropriate redshift-dependent sample
luminosity/magnitude limit $L_{\nu}^{min}$ (simply taken as the 
minimum observed luminosity of each sample at that redshift).  We compare these
predicted LFs to the corresponding observed
\citet{Boyle00} B-band and \citet{Miyaji00} soft X-ray LFs
at these redshifts.  The agreement between the observed and
predicted LFs is excellent at all observed
luminosities, and is reproduced for all low redshifts modeled. In \citet{H05b}, 
we have considered several approaches to calculate the column density 
distribution, and find that, after accounting for the clumping of most mass in 
some hot phase-cold phase separation, considering or ignoring metallicity and ionization 
results in factor $\sim2-3$ differences in the calculated column densities 
and corresponding quasar lifetimes along a given line of 
sight. Thus, we consider this to be a rough parameterization of the extremes of our modeling, 
and estimate 
the resulting range in the predicted luminosity functions as a 
consequence of 
these extremes
in calculations of the column density. We show the typical 
(averaged over the plotted points) resulting range in dex for both the 
optical and soft X-ray LFs at $z=1.0$ in the upper right of the corresponding 
panel, and note that the range at the other plotted redshifts is similar (though 
slightly smaller). The plotted range is significant in considering the relative 
luminosity functions, but we plot them not as errors but as an upper limit to the 
systematic effects of various extreme assumptions such as ignoring ionization and metallicity 
in calculating the column density and quasar attenuation. Based on our analysis above 
showing that near very peak luminosities the quasar will drive a wind expelling 
nearby gas and rapidly reducing the column density, we also consider 
a column density distribution which follows the above, fitted form for 
most of the quasar lifetime but cuts off (obscuration is neglected) when the 
quasar is within a fraction $\sim1/2$ of its peak luminosity. We find that this 
makes little difference to the predicted luminosity functions, primarily 
because the time spent at these very near peak luminosities is relatively small, 
and further because this results in a somewhat compensating small shift 
in the fitted $\nLP$ distribution. We note again though that a more complete 
treatment of this effect requires more complete spectral and dust modeling of 
both the quasar and stellar populations of the host galaxy, which we 
defer to a later paper. 
At higher redshift $z \gtrsim 1$, we recognize that the light curves and \NH\
distributions may evolve as a result of changing host galaxy
properties, and we defer a modeling of the LFs
at high redshifts to a future paper.

We also consider the results obtained using our column
density distributions, but applying only the idealized,
luminosity-independent models of the quasar lifetime described
above. The difference in lifetimes and the resulting $\nLP$ are
described in detail in \citet{H05c}, but essentially
$\nLP\propto\Phi(L=L_{peak})$ for these models.  For the simplest fit
to the \NH\ distributions in \S2, 
$\meanNH=\meanNH(L)\propto L^{0.35}$ and $\sigNH={\rm
constant}\approx0.8$, $f_{obs}$ is independent of $L_{peak}$ and can
be taken out of the integral, giving
$\Phi(\nu,L,L_{\nu}^{min})=f_{obs}(\nu,L)\,\Phi(L)$, independent of
the lifetime and $\nLP$ model.  However, even for a weak dependence on
$L_{peak}$, $\sigNH=\sigNH(L,L_{peak})=1.0+0.1\log{(L/L_{peak})}$, we
find that using these models of the quasar lifetime under-predict
both $\Phi_{B}$ and $\Phi_{SX}$ by a factor of $\gtrsim3$ at low and
high luminosities. This is because these models do not account
for the quasar spending most of its life at luminosities well below
its peak and thus do not properly account for quasars with different
$L_{peak}$ (i.e. different host galaxy properties such as total mass
or gas fraction) at a given observed luminosity. In any case, such a
procedure is not self-consistent, as the data from which our
$\meanNH(L,L_{peak})$ and $\sigNH(L,L_{peak})$ relations are fitted
imply and produce our model of luminosity-dependent quasar lifetimes,
with the vast majority of each $\nh(L)$ distribution corresponding to
points on the lightcurve which do not exist in these idealized
models.

\section{Conclusions\label{sec:conclusions}}

Using our picture of merger-driven quasar activity with
self-regulated black hole growth and feedback, we are able to {\em
simultaneously} reproduce the observed hard X-ray, soft X-ray, and
B-band luminosity functions (LFs) over a broad range of observed
luminosities and redshifts with significantly greater accuracy
than previous models and {\em without} invoking any assumptions beyond
the basic input physics of our simulations. Furthermore, our picture
also yields the observed $M_{BH}-\sigma$ relation \citep{DSH05}, the
bimodal distribution of galaxy colors \citep{SDH05a}, observed quasar
lifetimes \citep{H05a}, the \NH\ distribution of both optical and
X-ray samples \citep{H05b}, and the faint-end slope of the quasar
LF and supermassive black hole mass distribution
\citep{H05c}. We note that we are not predicting the luminosity function 
in this work, but are demonstrating that the observed differences between 
luminosity functions are essentially entirely accounted for in our modeling. In other 
words, given a luminosity function in any waveband, our modeling allows us to 
accurately determine 
the luminosity function in other wavebands, giving us robust 
predictive power across different samples and demonstrating that these 
observed differences can be explained self-consistently through a model 
in which observed patterns of quasar obscuration are dominated by differences in 
different stages of the {\em evolution} of quasars, 
and not simply by viewing-angle effects. 

Our model predicts that self-regulating feedback processes in galaxy
mergers reproduce the difference in the quasar LF at
different frequencies naturally, as a consequence of the evolution of
gas flows fueling accretion from gravitational torques
(e.g. Barnes \& Hernquist 1991, 1996; Mihos \& Hernquist 1996)
and accretion
feedback.  The population of obscured quasars is also a natural
consequence of the model, not as an independent population but as a
stage in the ``standard'' evolution of quasars over their lives,
before feedback can clear sufficient material to render the quasar
visible. Once feedback begins to unbind gas in the central regions of the 
galaxy, the quasar becomes observable in optical wavebands, and column densities 
rapidly decrease near the quasar peak luminosity. This suggests that our modeling of 
this blowout phase, coupled with the new interpretation of the luminosity 
function resulting from our model of quasar lifetimes, can explain the observed 
trends in the fraction of obscured or broad-line quasars with luminosity. 

The close agreement between our predictions and the observed
LFs is strong evidence in favor of our
self-consistent model of quasar lifetimes and light curves.  This
model suggests a new and qualitatively different interpretation of the
quasar LF, which we propose in \citet{H05c}.  Our
interpretation of the quasar LF and the intrinsic
deconvolved distribution of peak quasar luminosities and host galaxy
properties has important implications for the evolution of quasar
populations, the energetics of the cosmic X-ray, UV, and IR
backgrounds, the role played by quasars in reionization, and the
production of the present-day distribution of supermassive black
holes.  Future attempts to understand, model, or incorporate the
distribution of quasar properties should account for the difference
between the observed LF and the intrinsic
distribution of source properties as a result of luminosity dependent
quasar lifetimes, and the simultaneous, luminosity-dependent effects
of evolving \NH\ distributions. 

\acknowledgments Our thanks to the anonymous referee whose suggestions
greatly improved the quality and scope of this paper. 
This work was supported in part by NSF grants AST
02-06299, and AST 03-07690, and NASA ATP grants NAG5-12140,
NAG5-13292, and NAG5-13381.


\begin{thebibliography}{}
\bibitem[Alexander \etal(2005)]{Alexander05}
Alexander, D.~M., \etal\ 2005, Nature, 434, 738
\bibitem[Antonucci(1993)]{Antonucci93} 
Antonucci, R.\ 1993, \araa, 31, 473 
\bibitem[Barger \etal(2005)]{Barger05}
Barger, A.~J., Cowie, L.~L., Mushotzky, R.~F., Yang, Y., Wang, W.-H., Steffen, A.~T., \& Capak, 
P.\ 2005, \aj, 129, 578
\bibitem[Barnes \&\ Hernquist(1991)]{BH91}
Barnes, J. E. \&\ Hernquist, L. 1991, \apj, 370, L65
\bibitem[Barnes \&\ Hernquist(1996)]{BH96}
Barnes, J. E. \&\ Hernquist, L. 1996, \apj, 471, 115
\bibitem[Boroson(1992)]{Boronson92} 
Boroson, T.~A.\ 1992, \apjl, 399, L15 
\bibitem[Bouchet et al.(1985)]{Bouchet85} 
Bouchet, P., Lequeux, J., Maurice, E., Prevot, L., \&\ 
Prevot-Burnichon, M.~L.\ 1985, \aap, 149, 330 
\bibitem[Boyle et al.(1998)]{Boyle98} 
Boyle, B.~J., \etal\ 1998, \mnras, 296, 1 
\bibitem[Boyle et al.(2000)]{Boyle00} 
Boyle, B.~J., Shanks, T., Croom, S.~M., Smith, R.~J., Miller, L., 
Loaring, N., \& Heymans, C.\ 2000, \mnras, 317, 1014 
\bibitem[Busha et al.(2004)]{busha04}
Busha, M.T., Evrard, A.E., Adams, F.C. \& Wechsler, R.H. 2004, MNRAS,
submitted [astro-ph/0412161]
\bibitem[Comastri \etal(1995)]{Comastri95}
Comastri, A., Setti, G., Zamorani, G., \& Hasinger, G.\ 1995, \aap, 296, 1
\bibitem[Croom et al.(2004)]{Croom04} 
Croom, S.~M., Smith, 
R.~J., Boyle, B.~J., Shanks, T., Miller, L., Outram, P.~J., \& Loaring, 
N.~S.\ 2004, \mnras, 349, 1397 
\bibitem[Dav\'e et~al.(1999)]{dave99}
Dav\'e, R., Hernquist, L., Katz, N. \& Weinberg, D.H. 1999, \apj, 511, 521
\bibitem[Dawson \etal(2003)]{Dawson03}
Dawson, S., McCrady, N., Stern, D., Eckart, M.~E., Spinrad, H., Liu, M.~C., \&\ 
Graham, J.~R.\ 2003, \aj, 125, 1236
\bibitem[Di Matteo et al.(2005)]{DSH05}
Di Matteo, T., Springel, V., \&\ Hernquist, L. 2005, Nature, 433, 604
\bibitem[Elvis et al.(1994)]{Elvis94} 
Elvis, M., et al.\ 1994, \apjs, 95, 1
\bibitem[Elvis et al.(2002)]{ERZ02} 
Elvis, M., Risaliti, G., \& Zamorani, G.\ 2002, \apjl, 565, L75 
\bibitem[Fabian(1999)]{Fabian99}
Fabian, A.~C. 1999, MNRAS, 308, L39
\bibitem[George \etal(1998)]{George98}
George, I. M., Turner, T. J., Netzer, H., Nandra, K., Mushotzky, R. F., \&\ 
Yaqoob, T. 1998, \apjs, 114, 73
\bibitem[Gilli et al.(1999)]{Gilli99} 
Gilli, R., Risaliti, G., \& Salvati, M.\ 1999, \aap, 347, 424 
\bibitem[Gilli et al.(2001)]{Gilli01} 
Gilli, R., Salvati, M., \& Hasinger, G.\ 2001, \aap, 366, 407 
\bibitem[Green \etal(1995)]{Green95}
Green, P.~J., et al.\ 1995, \apj, 450, 51 
\bibitem[Grimes, Rawlings, \&\ Willott(2004)]{GRW04}
Grimes, J.~A., Rawlings, S., \&\ Willott, C.~J. 2004, \mnras, 349, 503
\bibitem[Hasinger(2004)]{Hasinger04}
Hasinger, G.\ 2004, Nucl.\ Phys.\ B Proc.\ Supp., 132, 86
\bibitem[Hernquist(1990)]{Hernquist90}
Hernquist, L. 1990, \apj, 356, 359
\bibitem[Hernquist(1993)]{H93}
Hernquist, L. 1993, \apj, 404, 717
\bibitem[Hopkins et al.(2004)]{Hopkins04} 
Hopkins, P.~F., et al.\ 2004, \aj, 128, 1112 
\bibitem[Hopkins \etal(2005a)]{H05a}
Hopkins, P.~F., Hernquist, L., Martini, P., Cox, T.~J., Robertson, B., Di Matteo, T., \&\ 
Springel, V. 2005a, \apj, submitted [astro-ph/0502241]
\bibitem[Hopkins \etal(2005b)]{H05b}
Hopkins, P.~F., Hernquist, L., Cox, T.~J., Di Matteo, T., Martini, P., \&\ 
Robertson, B., Springel, V. 2005b, \apj, submitted [astro-ph/0504190]
\bibitem[Hopkins \etal(2005c)]{H05c}
Hopkins, P.~F., Hernquist, L., Cox, T.~J., Di Matteo, T., \&\ 
Robertson, B., Springel, V. 2005c, \apj, submitted [astro-ph/0504252]
\bibitem[Katz et~al.(1996b)]{katz96b}
Katz, N., Weinberg, D.H. \& Hernquist, L. 1996b, \apjs, 105, 19
\bibitem[Kauffmann \& Haehnelt(2000)]{KH00} 
Kauffmann, G., \& Haehnelt, M.\ 2000, \mnras, 311, 576 
\bibitem[Kuraszkiewicz et al.(2000)]{Kuraszkiewicz00} 
Kuraszkiewicz, J., Wilkes, B.~J., Brandt, W.~N., \& Vestergaard, M.\ 2000, \apj, 542, 631
\bibitem[La Franca et al.(2002)]{LaFranca02} 
La Franca, F., et al.\ 2002, \apj, 570, 100 
\bibitem[Madau \etal(1994)]{Madau94}
Madau, P., Ghisellini, G., \& Fabian, A.~C.\ 1994, \mnras, 270, L17 
\bibitem[Magorrian et al.(1998)]{mag98}
Magorrian, J. et al. 1998, \aj, 115, 2285
\bibitem[Mainieri \etal(2005)]{Mainieri05}
Mainieri, V., \etal\ 2005, \aap, in press [astro-ph/0502542]
\bibitem[Marconi \etal(2004)]{Marconi04}
Marconi, A., Risaliti, G., Gilli, R., Hunt, L. K., Maiolino, R., \&\ Salvati, M. 
2004, \mnras, 351, 169
\bibitem[Martini(2004)]{Martini04}
Martini, P. 2004, in Carnegie Obs. Astrophys. Ser. 1, Coevolution of Black Holes
and Galaxies, ed. L.C. Ho (Cambridge: Cambridge Univ. Press), 170
\bibitem[McKee \&\ Ostriker(1977)]{MO77}
McKee, C. F. \&\ Ostriker, J. P. 1977, \apj, 218, 148
\bibitem[Mihos \&\ Hernquist(1996)]{MH96}
Mihos, J.~C. \&\ Hernquist, L.\ 1996, \apj, 464, 641
\bibitem[Miyaji et al.(2000)]{Miyaji00} 
Miyaji, T., Hasinger, G., \& Schmidt, M.\ 2000, \aap, 353, 25 
\bibitem[Morrison \&\ McCammon(1983)]{MM83}
Morrison, R. \&\ McCammon, D. 1983, \apj, 270, 119
\bibitem[{Navarro} et~al.(1996){Navarro}, {Frenk} \& {White}]{Navarro1996}
{Navarro} J.~F., {Frenk} C.~S., {White} S.~D.~M., 1996, \apj, 462, 563
\bibitem[Norman \etal(2002)]{Norman02}
Norman, C., \etal\ 2002, \apj, 571, 218
\bibitem[O'Shea et~al.(2005)]{oshea05}
O'Shea, B.W., Nagamine, K., Springel, V., Hernquist, L. \&
Norman, M.L. 2005, \apj, submitted
\bibitem[Page et al.(2004)]{Page04} 
Page, M.~J., Stevens, J.~A., Ivison, R.~J., \& Carrera, F.~J.\ 2004, \apjl, 611, L85 
\bibitem[Pei(1992)]{Pei92} 
Pei, Y. C. 1992, \apj, 395, 130
\bibitem[Perola \etal(2002)]{Perola02} 
Perola, G.~C., Matt, G., Cappi, M., Fiore, F., Guainazzi, M., 
Maraschi, L., Petrucci, P.~O., \&\ Piro, L.\ 2002, \aap, 389, 802
\bibitem[Richards \etal(2005)]{Richards05}
Richards, G.~T.\ \etal\ 2005, in press [astro-ph/0504300]
\bibitem[{Rines} et~al.(2004){Rines}, {Geller}, {Diaferio}, {Kurtz} \&
  {Jarrett}]{Rines2004}
{Rines} K., {Geller} M.~J., {Diaferio} A., {Kurtz} M.~J., {Jarrett} T.~H.,
  2004, \aj, 128, 1078
\bibitem[{Rines} et~al.(2002){Rines}, {Geller}, {Diaferio}, {Mahdavi}, {Mohr}
  \& {Wegner}]{Rines2002}
{Rines} K., {Geller} M.~J., {Diaferio} A., {Mahdavi} A., {Mohr} J.~J., {Wegner}
  G., 2002, \aj, 124, 1266
\bibitem[{Rines} et~al.(2000){Rines}, {Geller}, {Diaferio}, {Mohr} \&
  {Wegner}]{Rines2000}
{Rines} K., {Geller} M.~J., {Diaferio} A., {Mohr} J.~J., {Wegner} G.~A., 2000,
  \aj, 120, 2338
\bibitem[{Rines} et~al.(2003){Rines}, {Geller}, {Kurtz} \&
  {Diaferio}]{Rines2003}
{Rines} K., {Geller} M.~J., {Kurtz} M.~J., {Diaferio} A., 2003, \aj, 126, 2152
\bibitem[Robertson \etal(2004)]{Robertson04}
Robertson, B., Yoshida, N., Springel, V., \&\ Hernquist, L. 2004, \apj, 606, 32
\bibitem[Sazonov \& Revnivtsev(2004)]{sazrev04} 
Sazonov, S.~Y., \& Revnivtsev, M.~G.\ 2004, \aap, 423, 469
\bibitem[Schmidt(1968)]{Schmidt68}
Schmidt, M. 1968, ApJ, 151, 393 
\bibitem[Schmidt \&\ Green(1983)]{SG83}
Schmidt, M. \&\ Green, R.~F. 1983, ApJ, 269, 352 
\bibitem[Simpson(2005)]{Simpson05}
Simpson, C.\ 2005, \mnras, submitted [astro-ph/0503500]
\bibitem[Springel(2005)]{Spr05} 
Springel, V. 2005, \mnras, submitted [astro-ph/0505010]
\bibitem[Springel \&\ Hernquist(2002)]{SH02} 
Springel, V. \&\ Hernquist, L. 2002, \mnras, 333, 649
\bibitem[Springel \&\ Hernquist(2003)]{SH03} 
Springel, V. \&\ Hernquist, L. 2003, \mnras, 339, 289
\bibitem[Springel \etal(2005a)]{SDH05a}
Springel, V., Di Matteo, T., \&\ Hernquist, L. 2005a, \apj, in press, [astro-ph/0409436]
\bibitem[Springel \etal(2005b)]{SDH05b}
Springel, V., Di Matteo, T., \&\ Hernquist, L. 2005b, \mnras, submitted, [astro-ph/0411108]
\bibitem[Springel, Yoshida, \&\ White(2001)]{SYW01}
Springel, V., Yoshida, N., \&\ White, S. D. M. 2001, New Astronomy, 6, 79
\bibitem[Steffen \etal(2003)]{Steffen03}
Steffen, A.~T., Barger, A.~J., Cowie, L.~L., Mushotzky, R.~F., \&\ Yang, Y.\ 2003, ApJ, 596, L23
\bibitem[Stevens \etal(2005)]{Stevens05}
Stevens, J.~A., Page, M.~J., Ivison, R.~J., Carrera, F.~J., Mittaz, J.~P.~D., 
Smail, I., \&\ McHardy, I.~M.\ 2005, \mnras, submitted, [astro-ph/0503618]
\bibitem[Strateva \etal(2005)]{Strateva05}
Strateva, I., Brandt, N., Schneider, D.~P., Vanden Berk, D.~G., Vignali, C.\ 2005, \aj, in press [astro-ph/0503009]
\bibitem[Stern \etal(2002)]{Stern02}
Stern, D., et al.\ 2002, \apj, 568, 71
\bibitem[Telfer \etal(2002)]{Telfer02}
Telfer, R.~C., Zheng, W., Kriss, G.~A., \&\ Davidsen, A.~F.\ 2002, \apj, 565, 773
\bibitem[Tran(2003)]{Tran03} 
Tran, H.~D.\ 2003, \apj, 583, 632 
\bibitem[Treister \etal(2004)]{Treister04}
Treister, E., \etal\ 2004, \apj, 616, 123
\bibitem[Ueda et al.(2003)]{Ueda03} 
Ueda, Y., Akiyama, M., Ohta, K., \& Miyaji, T.\ 2003, \apj, 598, 886 
\bibitem[Vanden Berk \etal(2001)]{VB01} 
Vanden Berk, D.~E., \etal\ 2001, \aj, 122, 549 
\bibitem[Vignali \etal(2003)]{VBS03} 
Vignali, C., Brandt, W.~N., \&\ Schneider, D.~P.\ 2003, \aj, 125, 433
\bibitem[Volonteri et al.(2003)]{Volonteri03} 
Volonteri, M., Haardt, F., \& Madau, P.\ 2003, \apj, 582, 559 
\bibitem[Wilkes \etal(1994)]{Wilkes94}
Wilkes, B.~J., Tananbaum, H., Worrall, D.~M., Avni, Y., Oey, M.~S., \&\ Flanagan, J.\ 1994, \apjs, 92, 53 
\bibitem[Wyithe \& Loeb(2003)]{WL03} 
Wyithe, J.~S.~B., \& Loeb, A.\ 2003, \apj, 595, 614 
\end{thebibliography}
\end{document}